\documentclass[jap,aip,unsortedaddress,twocolumn,reprint]{revtex4-1}

\usepackage[latin1]{inputenc}
\usepackage{graphicx}
\usepackage{helvet}
\usepackage{times}
\usepackage{amsmath}
\usepackage{units}

\newcommand{\dd}{\mathrm{d}} 
\renewcommand{\vec}{\mathbf} 

\begin{document}

\title{Spatially resolved observation of uniform precession modes in spin-valve systems}

\author{Alexander M. Kaiser}\email{a.kaiser@fz-juelich.de}
\altaffiliation[Current address: ]{Materials Sciences Division, Lawrence Berkeley National Lab, Berkeley, CA 94720, USA.}
\author{Carsten Wiemann}%
 \author{Stefan Cramm}
\author{Claus M. Schneider}
\affiliation{Institut f\"ur Festk\"orperforschung IFF-9, Forschungszentrum J\"ulich and JARA-FIT, 52425 J\"ulich, Germany}%


\begin{abstract}
Using time-resolved photoemission electron microscopy the excitation of uniform precession modes in individual domains of a weakly coupled spin-valve system has been studied. A coupling dependence of the precession frequencies has been found that can be reasonably well understood on the basis of a macrospin model. By tuning the frequency of the excitation source the uniform precession modes are excited in a resonant way.
\end{abstract}

\pacs{68.37.Xy, 75.50.Bb, 75.78.-n, 76.50.+g, 78.20.Ls}
\keywords{Magnetization dynamics, PEEM, ferromagnetic resonance, interlayer coupling}
\maketitle

\section{Introduction}
The high-frequency dynamic response of magnetic thin film systems is an important research field regarding the operation of magnetic memory devices \cite{wolf}. The excitation of intrinsic precessional modes during the switching process can drastically effect the switching speed of the device \cite{schumacher2002csm, miltat}. In the simplest case, the dynamics of the magnetization $\vec{M}$ in the presence of a magnetic field $\vec{H}$ is given by the Landau-Lifshitz-equation\cite{landau1935tdm} $\nicefrac{\dd \vec{M}}{\dd \vec{H}} \propto - \vec{M} \times \vec{H}$ supplemented by the Gilbert term $\propto \alpha\left(\vec{M} \times \nicefrac{\dd \vec{M}}{\dd \vec{H}}\right)$ describing the damping\cite{gilbert2004ptd}. Disregarding inhomogeneities of field and magnetization, integration of that equation yields the Kittel frequency of uniform precession \cite{kittelfmr,miltat} $\omega_0 = \mu_0 \gamma \sqrt{(H_\text{eff} + M_S) \cdot H_\text{eff}}$. In reality, however, the magnetization $\vec{M}(\vec r)$ is inhomogeneous and the magnetic field is replaced by an effective field $\vec{H}_\text{eff}(\vec{r})$ composed of the external field and internal components, arising from anisotropies, magnetic coupling and demagnetizing fields \cite{magneticdomains}. In submicron sized samples the influence of such inhomogeneities becomes strong and hence inhomogeneous eigenmodes have to be considered such as precessional modes \cite{miltat, normalmodes}, domain wall modes \cite{parkspindynamics} or the gyrotropic motion of a vortex core \cite{choevortex}. 

The occurrence of such localized phenomena renders these systems an interesting playground for microscopy approaches allowing the study of the micromagnetic behavior directly. \textit{X-Ray photoemission electron microscopy} (X-PEEM) is one powerful method permitting magnetic investigations with lateral resolution \cite{schneiderschoenhense}. The magnetic sensitivity of this method is due to the \textit{X-Ray magnetic circular dichroism} (XMCD) \cite{schuetz, stoehrxmcdpeem}. By tuning the photon energy to adequate absorption edges element-selective and thus layer-selective studies can be performed. Combined with a stroboscopic illumination, time-resolved PEEM \cite{schoenhensetrpeem} has been used to study different types of excitation modes, such as modes of uniform precession \citep{krasyuk2005stm, raabe2005qam}, oscillations of the vortex core \citep{choevortex} and propagating and standing spin-waves \citep{wegelinspinwaves}. 

We have studied the magnetodynamic response of trilayers of Co$_{50}$Fe$_{50}$~(5~nm)/Cr/Ni$_{80}$Fe$_{20}$~(2~nm) grown on GaAs with a 200~nm thick Ag buffer layer. The system features a low magnetocrystalline anisotropy \cite{cofeanisotropy}, a weak oscillatory interlayer exchange coupling \cite{gruenbergiec} and different switching fields of the individual layers. The Cr layer has been grown as a wedge causing different coupling configurations on one sample \cite{UCP-91}. After the deposition, we defined the Ag coplanar waveguides with magnetic microstructures of several $\mu$m edge lengths by means of optical lithography and ion beam etching. The PEEM experiments have been carried out at the beamline UE56/1-SGM at BESSY-II (Berlin, Germany) at a light pulse repetition rate of 500~MHz. A function generator phase-locked to that frequency has been used to pass current pulses through the coplanar waveguides creating an magnetic field in the sample plane that acts on the magnetic microstructures. By varying the delay between the current pulses (pump) and the synchrotron light pulses (probe) the magnetodynamic evolution after the excitation of the magnetic field was studied.

\section{Pulsed excitation}

\begin{figure}
	\centering
		\includegraphics[width=\columnwidth]{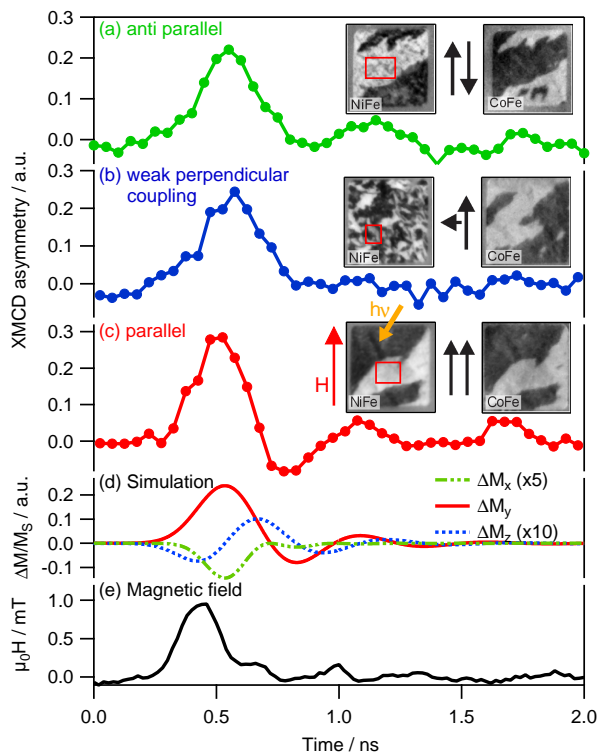}
	\caption{(a-c) Temporal profile of the response of the NiFe magnetization on the field pulse (d). The XMCD signal has been integrated over small areas (marked in the insets) within the magnetic structure yielding the highest response. The field and light incidence direction are marked by the red and orange arrows, respectively. All curves have been offset by the equilibrium XMCD value. (Inset) Equilibrium magnetization configurations of both layers for each configuration. (d) Macro-spin simulation of the three components of the free layer of a spin-valve system initially aligned parallel to $x$ and excited by a magnetic field pulse parallel to the $y$ direction. $M_x$ and $M_z$ are multiplied by a factor of 5 and 10, respectively.}
	\label{fig:timeresolved}
\end{figure}

Experiments with pulsed excitation have been conducted for three different square structures with an edge length of 12~$\mu$m at different positions along the spacer wedge corresponding to a different coupling strength between CoFe and NiFe. The resulting Cr thickness values were: \unit[1.7]{nm} (parallel coupling), \unit[2.0]{nm} (transition regime, partly 90° coupling) and \unit[2.3]{nm} (antiparallel coupling). The equilibrium domain structures are displayed in the insets of Fig.~\ref{fig:timeresolved}. 

Under the influence of a 250~ps long and 1~mT strong magnetic field pulse (shown in Fig.~\ref{fig:timeresolved} (e)) the system behaves as a pseudo spin-valve with a sole excitation of the NiFe magnetization and the CoFe layer being unaffected. Fig.~\ref{fig:timeresolved} (a-c) compiles the XMCD signals of the NiFe layer in all three microstructures. In each case, the XMCD signal has been integrated over areas with maximum response (marked in the insets). We observe a rotation of the NiFe magnetization towards the magnetic field direction returning shortly after the field pulse has decayed. The response peak is shifted by  $(100 \pm 50)~\text{ps}$ with respect to the pulse peak. In the case of parallel and antiparallel coupling configurations we observe a damped oscillation around the equilibrium position after the initial deflection. In the reaction of the weakly coupled structure only the direct rotation by the field pulse shows up. Domain wall motion is not observed.


The observed behavior can be understood as the excitation of a damped uniform precession in the individual domains. According to the Landau-Lifshitz term, the excitation is highest for a perpendicular orientation of $\vec{M}$ and $\vec{H}_\text{eff}$. After the initial excitation $\vec{M}$ and $\vec{H}_\text{eff}$ are not parallel anymore. Thus, the NiFe magnetization precesses around the effective field direction. The Gilbert term in turn leads to a decrease of the precessional amplitude and an alignment towards the direction of $\vec{H}_\text{eff}$. The $\vec{M} \times \vec{H}$-term also explains the delay between the observed rotation of $\vec{M}$ and the field peak. The field induces a small rotation of the magnetization perpendicular to $\vec{M}$ and $\vec{H}$, i.e. normal to the sample plane. After the decay of the magnetic field pulse, the internal effective field forces the magnetic moments back to their equilibrium position on a spiral trajectory. As our measurements are sensitive to changes parallel to the beam direction, we do not observe the first rotation out of the sample plane.

Neglecting the low magnetocrystalline anisotropy of NiFe, the main components contributing to $\vec{H}_\text{eff}$ are the demagnetizing field $\vec{H}_\text{demag} = -\vec{\mathsf{N}} \vec{Ms}$ with the demagnetizing tensor $\vec{\mathsf{N}}$ and the coupling field $H_\text{coupl} = J/(\mu_0 M_s\cdot t)$ with the coupling constant $J$ and the film thickness $t$. From fits of a damped sine function to the measured data, frequencies of $f_\text{parallel} = (1.62 \pm 0.14)~\text{GHz}$ (parallel) and $f_\text{antiparallel} = (1.7 \pm 0.1)~\text{GHz}$ with damping times of $\tau_\text{parallel} = (0.23 \pm 0.08)~\text{ns}$ and $\tau_\text{antiparallel} = (0.25 \pm 0.1)~\text{ns}$ may be deduced. Using the above mentioned relations and a magnetization of NiFe of $\mu_0 M_s = \unit[1]{T}$\cite{cofeanisotropy}, the effective fields and damping constants have been calculated as $\mu_0 H_{\text{eff, P}} = (3.3 \pm 0.3)~\text{mT}$ and $\alpha_\text{P} = 0.048 \pm 0.017$ for the parallel coupling and $\mu_0 H_{\text{eff, AP}} = (3.4 \pm 0.2)~\text{mT}$ and $\alpha_\text{AP} = 0.044 \pm 0.009$ for the antiparallel coupling, respectively. These effective field values are much larger than the estimated demagnetizing field of the order of $\mu_0 H_\text{demag} \approx \unit[0.4]{mT}$\cite{aharoni}. Subtracting this value, a coupling field of $\mu_0 H_\text{coupl} \approx \unit[3]{mT}$ can be derived, from which a coupling constant $|J| \approx \unit[5\cdot 10^{-6}]{J/m^2}$ is obtained, in good agreement with the assumption of a weakly coupled system. The obtained result of the damping constant of $\alpha > 0.04$ is distinctly higher than the literature value of $\alpha = 0.01$ \citep{cofeanisotropy}. Similarly increased values in thin NiFe layers in heterostructures have also been found by other authors \citep{wegelinspinwaves, schumacher2002csm} and may be attributed to a spin-pumping effect due to the interaction with the neighboring non-magnetic films \citep{tserkovnyak2002egd}.


In order to further interpret the experimental results, the Landau-Lifshitz-Gilbert equation (LLG) was solved for a system of two coupled macrospins\cite{miltat} with the same dimensions as the magnetic systems investigated. The assumption of a macrospin is valid in the case of the collinearly coupled systems because then the inhomogeneous $H_\text{demag}$ is distinctly smaller than $H_\text{coupl}$ which can be assumed uniform throughout one domain. In the simulation one of the layers was fixed parallel to the $x$-axis and a weak Gaussian magnetic field pulse with an amplitude of $\unit[1]{mT}$ and a width of $\unit[250]{ps}$ was applied along the $y$-axis \cite{Note1}. The resulting response of the three components of $\vec{M}$ are shown in Fig.~\ref{fig:timeresolved} (d). The temporal characteristics for parallel ($J>0$) and antiparallel ($J < 0$) coupling are well reproduced by the simulation of the $M_y$ component and is not changed upon change of sign of $J$. Furthermore, the simulation reproduces the delay between the maximum deflection of the $M_y$ component and the field peak. A phase shift of $\unit[120]{ps}$ is obtained in good agreement with the experimental value of $(100 \pm 50)\unit{ps}$. As expected, the maximum $M_z$ deflection coincides with the maximum of the external field and is transformed with a delay in to an in-plane rotation. 

Thus, we conclude that in the case of parallel and antiparallel coupling the magnetization precesses around an effective field that is mostly composed of the coupling field favoring a parallel/antiparallel alignment of the NiFe magnetization to the magnetization of the magnetically harder CoFe layer. In both cases, the coupling constant is found to take a value of $|J| \approx \unit[5\cdot 10^{-6}]{J/m^2}$ but positive (parallel coupling) or negative (antiparallel) sign. The absence of such precessional motion in the case of weak coupling is attributed to the lower effective field with stronger inhomogeneous components blocking the magnetization rotation.

\section{Resonant excitation}
\begin{figure}
	\centering
		\includegraphics[width=1\linewidth]{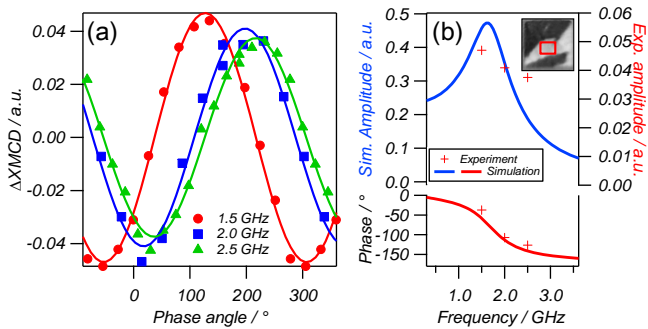}
	\caption{(a) Phase-resolved XMCD signals acquired with excitation frequencies of 1.5~GHz, 2.0~GHz and 2.5~GHz together with a sine fit. The signals have been integrated over the box marked in the inset and normalized to the excitation amplitude measured from the image shift. (b) Theoretical frequency dependence of oscillation amplitude and phase as acquired from a macrospin simulation. Experimental values are marked with crosses.}
	\label{fig:phasesim}
\end{figure}

Due to their broad frequency spectrum ultrashort magnetic field pulses can excite a large spectrum of different oscillatory modes \citep{adam2007dpa, normalmodes, parkspindynamics}. By using a continuous RF sine wave with a single frequency, the excitation can be limited to selected modes. In combination with a spatially resolving technique this resonant excitation allows one to study the lateral distribution of such eigenmodes. This technique has also been referred to as \textit{spatially-resolved ferromagnetic resonance} (SR-FMR) \citep{tamaru2002iqm, puzic2005srf}. We have applied a similar approach to the investigation of magnetization precession with PEEM by using a continuous sine wave excitation with variable frequency, which was set to an integer multiple of the light pulse repetition frequency of \unit[500]{MHz} in order to retain the synchronization between pump and probe.

Experimental results on a structure with parallel coupling have been obtained by exciting the systems at frequencies of 1.5~GHz, 2.0~GHz and 2.5~GHz. The XMCD signals have been integrated over a large domain with $\vec{M}$ perpendicular to $\vec{H}$ and are plotted in Fig.~\ref{fig:phasesim} against the phase angle relative to the excitation signal. The coplanar waveguides and the cabling used in the experiment have a frequency-dependent transmission in the GHz-range. Thus, the magnetic field amplitudes have been measured directly from the shift and breathing of the PEEM images and the XMCD curves have been normalized to the so-obtained field amplitudes. The measured signals can be approximated by sine fits with different amplitude and phase values. The phase is strongly frequency-dependent while the normalized amplitude changes only slightly. 

Macrospin simulations using the same simulation parameter as described above have been conducted assuming an RF excitation with variable frequency and the resulting amplitude and phase values of the $M_y$ rotation are displayed in Fig.~\ref{fig:phasesim} (b). The simulations predict a resonance at a frequency of \unit[1.6]{GHz} connected with an increase of the oscillation amplitude and a phase shift of 180° over the width of the resonance ($\Delta f \approx \unit[1]{GHz}$). The experimental phase values show the same characteristics as the calculated values. However, the simulated behavior of the amplitudes cannot be reproduced quantitatively. This deviation may be ascribed to the differences in the applied field strength and the difficulties in determining the real amplitudes. Furthermore, shortcomings of the macrospin model due to the inhomogeneous constituents of $H_\text{eff}$ may play a role. However, as the general shape of the resonance curve is reproduced, the results evidence the existence of a uniform precession mode with an eigenfrequency of 1.6~GHz as deduced from the pulsed excitation experiments.

\section{Summary \& Outlook}

In summary, we have used time-resolved PEEM for the study of uniform precession modes in single domains of a weakly coupled spin-valve system. The results revealed a coupling-dependent precession frequency. By developing an experimental approach of frequency-dependent resonant excitation, we have been able to study the precession at different frequencies revealing a frequency-dependency of amplitude and phase of the magnetodynamic response. All experimental results were shown to be in good agreement with theoretical predictions of a macrospin model. The frequency-dependent experiments reproduced the shape of a resonance curve with the resonance frequency as derived from the experiments using pulsed excitation proving the feasibiliy of spatially-resolved FMR experiments employing PEEM. In future experiments the approach of resonant excitation may be used for systematic studies of particular localized eigenmodes in magnetic systems.

\acknowledgements
We would like to thank R. Schreiber for the sample preparation and K. Bickmann, J. Lauer, B. Küpper and H. Pfeifer for their technical support. This work has been financially supported by the DFG (SFB 491) and the BMBF (Project 05KS7UK1).

\bibliographystyle{aip}


\end{document}